\documentclass[twocolumn,secnumarabic,amssymb, nobibnotes, aps, prd]{revtex4-1}
\pdfoutput=1 
\usepackage[T1]{fontenc} 
\usepackage[utf8]{inputenc}
\usepackage{amsfonts} 
\usepackage{graphics}
\usepackage{epsfig} 
\usepackage[english]{babel} 
\usepackage{amsmath}
\usepackage{bm}
\usepackage{multirow}
\usepackage{array,booktabs}
\usepackage{calc}

\begin{document}

%
\title{Possibility of observing Higgs bosons at the ILC in the lepton-specific 2HDM}
\author{Majid Hashemi}
\email{majid.hashemi@cern.ch} 
\affiliation{Physics Department, College of Sciences, Shiraz University, \\ Shiraz, 71946-84795, Iran}

%
\begin{abstract}
The Higgs boson pair production at a linear $e^+e^-$ collider is analyzed in the $4\tau$ final state in the context of lepton-specific or type IV 2HDMs. Both beams are assumed to be unpolarized. The Higgs boson pairs ($HA$) are produced through off-shell $Z^*$  production and decay to $\tau$-jets which is the main decay channel for neutral Higgs bosons in 2HDM type IV. Using a simplified detector simulation based on the SiD detector at ILC, the $4\tau$ signal is studied through the $\tau$-jet pair invariant mass reconstruction. Several benchmark scenarios are considered for center of mass energies of 500 GeV and 1000 GeV at integrated luminosity of 500$fb^{-1}$. Among Standard Model (SM) background processes, the main background is $e^+e^- \rightarrow ZZ$ followed by $Z \rightarrow \tau\tau$. This background is however, well under control. With the luminosity assumed in the analysis, striking signals are obtained beyond the reach of LHC. Such signals would allow for precise determination of masses and cross sections and already much lower luminosities are sufficient for discovery.
\end{abstract}
\maketitle
\section{Introduction}
After the discovery of the Higgs boson at the LHC  \cite{HiggsObservationCMS,HiggsObservationATLAS} which was predicted through a theoretical framework known as the Higgs mechanism  \cite{Englert1,Higgs1,Higgs2,Kibble1,Higgs3,Kibble2}, attention has been paid to the question whether the observed particle belongs to a single SU(2) doublet or is part of an extended structure such as a two Higgs doublet model (2HDM)  \cite{2hdm1,2hdm2,2hdm3}. Since the latter scenario can be made in a way to provide a light Higgs boson which respects the observed particle properties, one may expect an SM like structure consistent with experimental data plus new particles arising from the extended Higgs sector. Such additional Higgs particles can in general be different from the observed particle in terms of their masses and their couplings with SM fermions. Therefore one way to observe such particles would be to benefit from their characteristic features and decay channels which are different from those of the SM Higgs boson.  

The additional Higgs bosons of such a model are assumed to be heavier than the observed one. Therefore, a center-of-mass energy above the threshold of their masses is required to observe them. 

Obviously LHC is able to provide the effective center-of-mass energy required to produce heavy 2HDM Higgs bosons, however, in recent studies we have shown that the ability of linear colliders like ILC is much beyond LHC in observing their signals with a high statistical significance. In \cite{H24b,H24b2l} it was shown that signals from the type I 2HDM Higgs bosons can well be observed at $e^+e^-$ colliders through $H/A \rightarrow b\bar{b}$. The leptonic decay of the type IV 2HDM Higgs bosons through $H/A \rightarrow \mu\mu$ was also shown to provide clear signals on top of the background \cite{Hashemi2017,Hashemi2017_2}. Results of the current analysis are also beyond the reach of LHC. The main reasons for such successful results can be summarized as follows. 

First, the $e^+e^-$ collisions provide a cleaner environment in terms of less particle multiplicity and hadron activity. 

Second, some SM processes are simply absent in $e^+e^-$ colliders because of the electric charge conservation. These processes include single $W^{\pm}$ boson production and $W^{\pm}Z$ pair production. 

Third, the SM background processes have a smaller cross section at $e^+e^-$ colliders. As an example, while the top quark pair production, $t\bar{t}$, acquires a high cross section of $~800~pb$ at LHC, it can appear through off-shell $Z^*$ boson decay at $e^+e^-$ colliders with a cross section of less than $1~pb$ at $\sqrt{s}=500$ GeV. 

The above arguments are not the only ones but can be considered as the main features which discriminate $e^+e^-$ colliders from LHC. 
 
A general 2HDM may be categorized into four CP-conserving types with different scenarios of Higgs-fermion couplings. The ratio of vacuum expectation values of the two Higgs doublets ($\tan\beta=v_2/v_1$) is the free parameter of the model and leads to enhancement or suppression of Higgs-fermion couplings compared to the corresponding SM couplings \cite{tanbsignificance}.

In total five physical Higgs bosons are predicted in 2HDM. The lightest Higgs boson, $h$, (sometimes denoted as $h_{SM}$) is assumed to be the SM-like Higgs boson with the same couplings with fermions as in SM. There are two heavier neutral Higgs bosons, $H$ (CP-even) and $A$ (CP-odd), and two charged Higgs bosons, $H^{\pm}$ \cite{2hdm_TheoryPheno}.

The search for the 2HDM Higgs bosons has been one of the main programs of high energy physics experiments since the time of LEP. 

In what follows a brief review of the relevant experimental studies are presented. The relevance is in terms of either the 2HDM type IV or the final states involving $H/A \to \tau\tau$ decay. 







One of the first analyses of the $\tau\tau$ final state is from DELPHI collaboration with $e^+e^- \to hZ$ and $hA$ as the signal followed by $h/A \to b\bar{b}, \tau\tau$ at $\sqrt{s}=189-208$ GeV. Their results were presented as limits on the branching ratio of the Higgs boson decay to $b\bar{b}$ and $\tau\tau$ \cite{Abdallah:2004wy}.


The CDF collaboration reported analysis of $p\bar{p} \to HA$ with $H/A \to \tau\tau$ using 1.8 $fb^{-1}$ of integrated luminosity and excluded $\tan\beta > 50$ for a Higgs boson mass range $90 < m_A < 180$ GeV \cite{Aaltonen:2009vf}. 

The D0 collaboration also published results of several studies. In a search for associated production of a Higgs boson and a $b$ quark, followed 
by $H/A \to \tau\tau$,~ $\tan\beta > 30$ was excluded for $90 <m_A < 180$ GeV \cite{Abazov:2011qz}. 

With more data using $H/A \to \tau\tau$ decays, the most stringent result of $p\bar{p}$ collisions was obtained by excluding $\tan\beta > 20$ for Higgs boson mass range $90 <m_A < 180$ GeV \cite{Abazov:2011up}.  

The CMS and ATLAS collaborations reported their analyses of $A \to Zh$ decay in the $\ell\bar{\ell}b\bar{b}$ final state based on $m_A=300$ GeV assumption and presented exclusion contours for 2HDM types I and II (CMS) \cite{Khachatryan:2015lba} and all four types (ATLAS) \cite{Aad:2015wra}. The lepton-specific 2HDM interpretation of the results by the ATLAS collaboration excluded $\tan\beta < 10$ \cite{Aad:2015wra}. One of the most recent analyses of all 2HDM types by the ATLAS collaboration has been presented based on a search for $A \to ZH$ decay at low $\tan\beta$ \cite{Aaboud:2018eoy}. The $A \to ZH$ decay occurs when $m_A-m_H$ reaches the $Z$ boson mass threshold and prefers low $\tan\beta$ because at high $\tan\beta$ the leptonic decay grows. The exclusion contours obtained in \cite{Aaboud:2018eoy} are thus shown for $\tan\beta<3$ and $m_A-m_H>90$ GeV. 



The above reports are based on real data searches, while there are studies carried out for ILC and CLIC environment in the literature.  

The Higgs boson program of the ILC begins at $\sqrt{s}=250$ GeV which is near the peak of the cross section for $e^+e^- \to Zh$ \cite{ilctdrv4}. One of the main operating scenarios of ILC is to perform $e^+e^-$ collisions at $\sqrt{s}=500$ GeV \cite{ilc_operating_scenarios}. Ultimately, the ILC is upgradable to a center-of-mass energy of 1 TeV \cite{ilctdrv4}. 

The Higgs physics at CLIC has also been extensively studied and presented in the CLIC conceptual design report \cite{cliccdr1,cliccdr2,cliccdr3}. In a recent review, a detailed study of different production processes and decay channels of the SM Higgs boson has been preseneted in \cite{HiggsCLIC}. 

\begin{figure}[h]
\centering  \includegraphics[width=0.45\textwidth,height=0.4\textwidth]{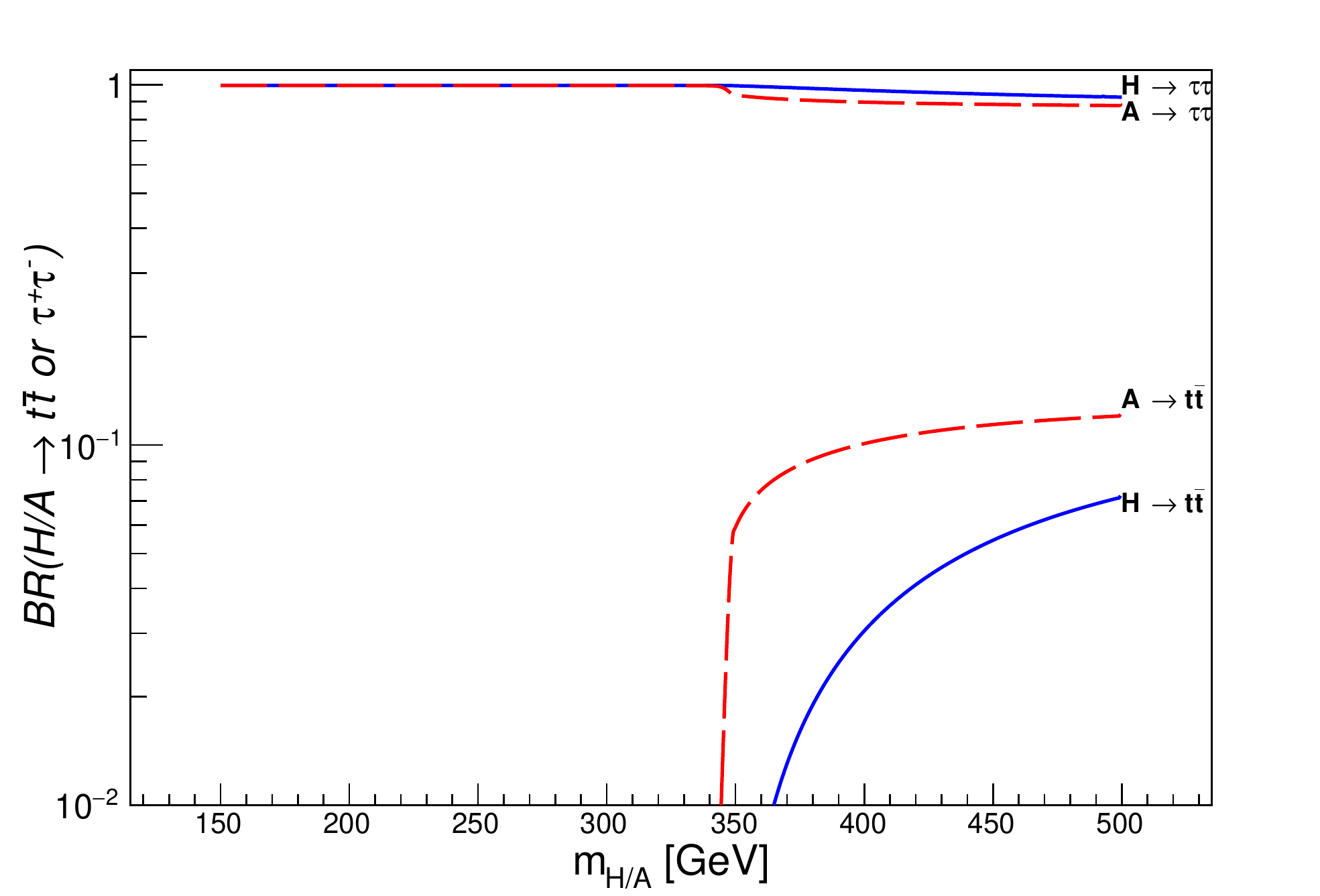}
  \caption{BR(H/A $\to$ XY) as a function of the Higgs bosons masses \label{BRs}}
  \end{figure}

\begin{figure}[h]
\centering  \includegraphics[width=0.45\textwidth,height=0.4\textwidth]{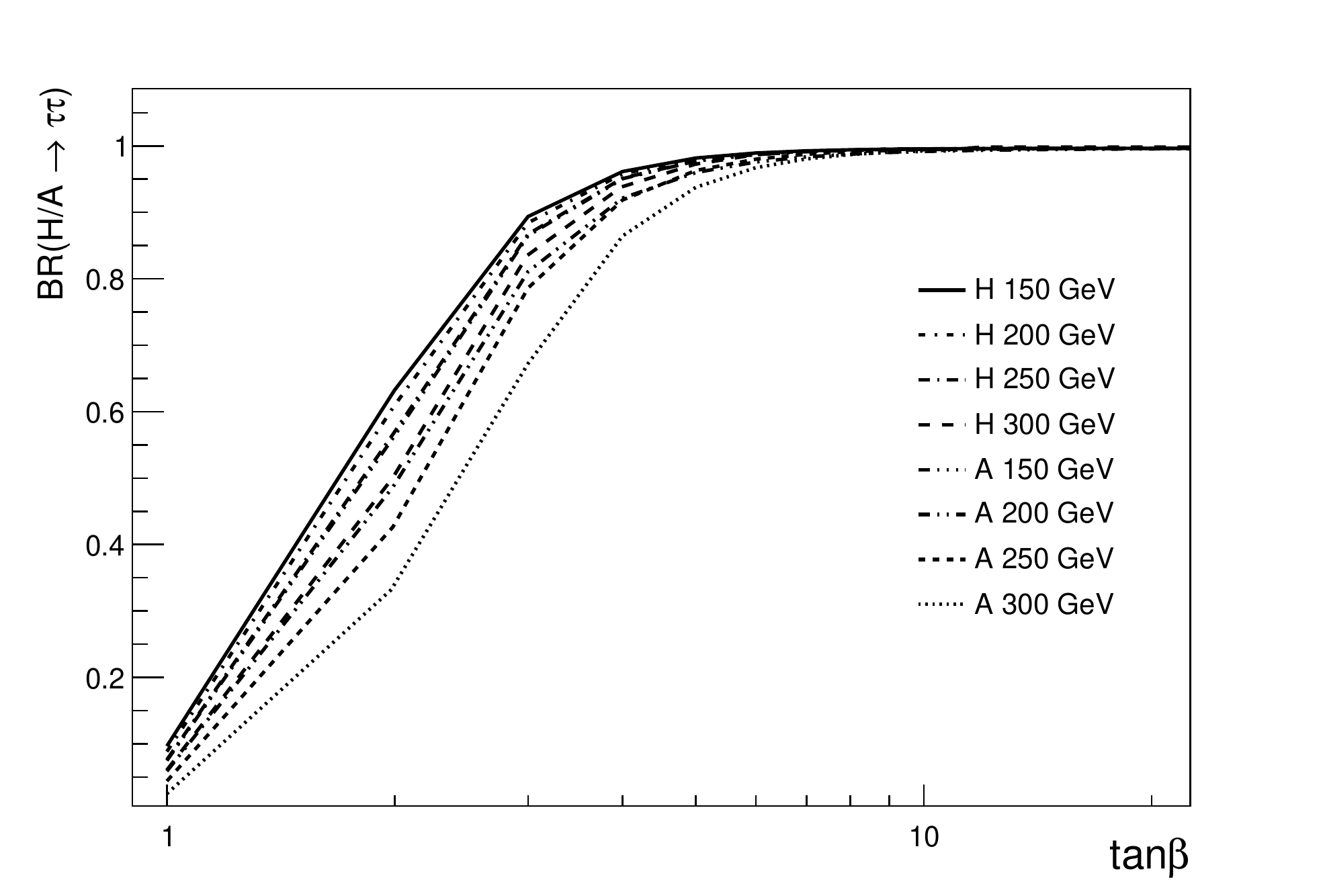}
  \caption{BR(H/A $\to$ XY) as a function of $\tan\beta$ \label{BR_vs_tb}. The selected points showing $H$ decays assume $m_A=150,~250,~250,~300$ GeV from top to bottom. Similarly, points showing $A$ decays assume $m_H=150,~150,~250,~300$ GeV.}
  \end{figure}

The multi-$\tau$ final state of Higgs boson pair production has been shown to be a promising channel at LHC \cite{Kanemura1} and ILC \cite{Kanemura2}. Using a proper $\tau$ tagging algorithm, the four $\tau$ jets can be identified and used to reconstruct the Higgs bosons masses. There are enough kinematic constraints to reconstruct the two Higgs bosons with a high accuracy \cite{MSSMHiggsLEP, Kanemura2}. This idea is used in the current analysis. The focus is on 2HDM type IV which allows for the heavy neutral Higgs boson decay to leptons while suppressing other decay channels. Since the $\tau$ lepton is the heaviest lepton, the decay to a $\tau$ pair is the dominant channel because the relevant Higgs-lepton couplings depend on the lepton mass. As Fig. \ref{BR_vs_tb} shows, at $\tan\beta=3$, the Higgs bosons decay to $\tau\tau$ reaches $\sim 80\%$. Therefore below the top quark pair production threshold, $H/A \rightarrow \tau\bar{\tau}$ is dominant, while above the threshold there is small reduction in the $H/A \rightarrow \tau\tau$ decay allowing $H/A \rightarrow t\bar{t}$ to appear at the level of a few percent. As seen from Fig. \ref{BRs}, $H/A \rightarrow t\bar{t}$ reaches 10$\%$ at high masses leading to $\sim 10 \%$ reduction of $H/A \rightarrow \tau\tau$ around $m_{H/A}\simeq 500$ GeV. Since both neutral Higgs bosons (H and A) decay to $\tau$ lepton pairs, the signal process has a 4$\tau$ signature to be distinguished from SM background processes like $e^+e^- \rightarrow ZZ \rightarrow 4\tau$.  

The 2HDM type IV receives soft limits from flavour physics studies at low $\tan\beta$ \cite{Misiak,Misiak2017,FM}. Therefore the region of study is wide in $\tan\beta$ direction starting from values as low as $\tan\beta \simeq 1$ to 50. The signal process, i.e., $e^{+}e^{-} \rightarrow Z^{(*)} \rightarrow HA$ is independent of $\tan\beta$ as the $ZHA$ vertex does not depend on Higgs-fermion couplings and the 2HDM type. Therefore the signal process including Higgs bosons decays is effectively independent of $\tan\beta$. This is a dramatic feature of the signal under study as it makes it independent of any parameter other than the center of mass energy of the collider and the Higgs bosons masses. Therefore only kinematic effects can change the signal cross section and its observation chance.

The strategy of the analysis is to generate signal and backgroud events and apply ILC detector simulation and perform $\tau$ identification algorithm using hadronic final state of $\tau$ leptons. The invariant mass of the two closest $\tau$ jets is then calculated and fills a distribution which serves as the Higgs boson candidate invariant mass. The same approach is applied on background events and a final assessment is made on the possibility of signal observation using statistical techniques. Before going to the details of the analysis, a brief review of the theoretical framework is presented in the next section.

\section{Theoretical framework}
The Higgs-fermion couplings in a general 2HDM appear as a Yukawa Lagrangian as in Eq. \ref{lag} \cite{2hdm_HiggsSector1}.
\begin{align}
\begin{split}
\mathcal{L}_Y&=\sum_{f=U,D,L}\left[\rho^{f}s_{\beta-\alpha}-\kappa^f c_{\beta-\alpha}\right]\overline{f}fH \\
&-i \sum_{f=U,D,L}\rho^{f}_A\overline{f}\gamma_5 fA\\
\end{split}
\label{lag}
\end{align}
in which $U,D,L$ are the up-type and down-type quarks and leptons fields, $H$ and $A$ the neutral Higgs boson fields, $\kappa^f=\frac{m_f}{v}$ are the SM Higgs-fermion couplings, $s_{\beta-\alpha}=\sin(\beta-\alpha)$ and $c_{\beta-\alpha}=\cos(\beta-\alpha)$. Here $\alpha$ is the neutral Higgs mixing angle. 

The $\rho^f$ parameters depend on the 2HDM type and are proportional to $\kappa^f$ as shown in Tab. \ref{typeIV} specifically for type IV \cite{Barger_2hdmTypes}. The CP-odd Higgs couplings ($\rho^f_A$) are the same as $\rho^f$ except for an additional minus sign for $f=U$. Therefore the neutral CP-even Higgs couplings depend on the values of $\rho^f$ which are $\kappa^f$ (as in SM) times a $\tan\beta$ or $\cot\beta$ factor which leads to possible deviations from SM \cite{2hdm_HiggsSector2}.

\begin{table}[h]
\centering
\begin{tabular}{ccc}
\midrule
\multicolumn{3}{c}{2HDM Type IV}\\
\midrule
		 $\rho^D$ & $\rho^U$ & $\rho^L$ \\
\midrule
$~\kappa^D \cot\beta~$  & $~\kappa^U \cot\beta~$  & $~-\kappa^L \tan\beta~$  \\
\midrule
\end{tabular}
\caption{The type IV Higgs boson couplings with $U$(up-type quarks), $D$(down-type quarks) and $L$(leptons).\label{typeIV}}
\end{table}

The light Higgs boson of 2HDM behaves the same as the SM Higgs by setting $s_{\beta-\alpha}=1$ (the SM-like limit). This ensures the same couplings of the light Higgs with fermions and gauge bosons as in SM. Therefore one can assume that the observed particle is the light Higgs boson of 2HDM. On the other hand, the above setting suppresses the heavy neutral CP-even Higgs coupling with gauge bosons which is proportional to $c_{\alpha-\beta}$ \cite{2hdm_TheoryPheno}. 

Under the assumption $s_{\beta-\alpha}=1$, the brief form of the Lagrangian takes the form:
\begin{align}
\begin{split}
\mathcal{L}_Y&=\sum_{f=U,D,L}\rho^{f}\overline{f}fH \\
&-i \sum_{f=U,D,L}\rho^{f}_A\overline{f}\gamma_5 fA\\
\end{split}
\label{lag2}
\end{align}
which is translated into the expanded explicit mode of Eq. \ref{lag3} when Tab. \ref{typeIV} is used.
\begin{align}
\begin{split}
\mathcal{L}_Y&=\frac{m^D}{v}\cot\beta\overline{D}DH+\frac{m^U}{v}\cot\beta\overline{U}UH\\
&-i \frac{m^D}{v}\cot\beta\overline{D}\gamma_5 DA+i\frac{m^U}{v}\cot\beta\overline{U}\gamma_5 UA\\
&+\frac{m^L}{v}\tan\beta\overline{L}LH-i \frac{m^L}{v}\tan\beta\overline{L}\gamma_5 LA\\
\end{split}
\label{lag3}
\end{align}

In such a scenario, Higgs boson conversion through $A \rightarrow ZH$ is also suppressed because at high $\tan\beta$ the leptonic decay dominates over the $\tan\beta$ independent $A \rightarrow ZH$ decay. Therefore both Higgs bosons, $H$ and $A$, decay to $\tau$ pairs and the signal (invariant mass distribution of the two $\tau$ jets) contains $\tau$ jet pairs from both Higgs bosons. 

It should be noted that a slight deviation from the SM-like limit (i.e., $s_{\beta-\alpha}\ne 1$) may drastically change the production and decay rates of the Higgs bosons \cite{Kanemura4}. 

The signal process in this analysis, i.e., $e^+e^- \to Z^* \to HA \to 4\tau$ depends on $s_{\beta-\alpha}$ through the $H.A.Z$ vertex and decreases if $s_{\beta-\alpha}\ne 1$. Under the same assumption, the fermionic decay $H/A \to \tau\tau$ dicreases due to the possibility of $A \to Zh$ \cite{Khachatryan:2015lba, Aad:2015wra}, $H \to hh$ \cite{Khachatryan:2015tha} and $H \to VV$ ($V=W, ~Z$) \cite{Aad:2015kna}. 

Detecting patterns of deviations in the SM-like Higgs boson coupling constants with precision data can fingerprint extended Higgs sectors. In a detailed study, expected precision of the Higgs boson coupling with gauge bosons has been obtained to be at the level of $4-6\%$ at LHC high luminosity run at $\sqrt{s}=14$ TeV and $0.39-0.49\%$ at ILC operating at $\sqrt{s}=500$ GeV \cite{Kanemura3}. The extended Higgs sectors can thus be indirectly tested by such precision measurements as the pattern of the deviations strongly depends on the structure of the Higgs sector (in this case, the type of 2HDM). 

In the case of Higgs-gauge coupling, a precision measurement with 0.5 $\%$ uncertainty can be used to set upper limits on the value of $\alpha$ parameter as the Higgs-gauge coupling in 2HDM has an extra factor of $\sin(\beta-\alpha)$ compared to the corresponding coupling in SM. If it truns out that the coupling measurement is compatible with SM ($\sin(\beta-\alpha)=1$) with 0.5 $\%$ uncertainty, then $0.99<\sin(\beta-\alpha)<1$ at 2$\sigma$ (95$\%$ C.L.). For example, for $\tan\beta=20$, this would leave an allowed region of $-0.05 < \alpha < 0.09$.
\section{Signal identification and the search scenario}

The signal process is chosen to be $e^{+}e^{-} \rightarrow Z^{(*)} \rightarrow HA \rightarrow 4\tau$. Only the hadronic decay of the $\tau$ lepton is considered to benefit from the unique features of $\tau$ jets in the detector. The center of mass energy of the collider should be high enough to produce both Higgs bosons. Here two scenarios of $\sqrt{s}=500$ and 1000 GeV are considered. 

The analysis is based on two sets of benchmark points which are selected separately for each center-of-mass energy. Equal and different Higgs boson masses are included.

All benchmark points are consistent with the theoretical requirements including potential stability, perturbativity and unitarity as checked by using \texttt{2HDMC 1.7.0} \cite{2hdmc1,2hdmc2}. The selected points are also consistent with experimental limits according to \texttt{HiggsBounds} \cite{HB} and \texttt{HiggsSignal} \cite{HS} and results of all new references mentioned in the introduction.

For each selected point in the parameter space, an \texttt{SLHA} file \cite{slha1, slha2} (containing particles mass and decay spectrum) is provided by \texttt{2HDMC} and is passed to \texttt{PYTHIA 8.2.15} \cite{pythia81,pythia82} for event generation and cross section calculation.  

Both beams are assumed to be unpolarized in the current analysis, although there are plans for polarized beam particles with polarization fraction of $80\%~(30\%)$ for electrons (positrons) at $\sqrt{s}=500$ GeV and $80\%~(20\%)$ at $\sqrt{s}=1000$ GeV \cite{ilctdrv4}. 

\begin{figure}[h]
    \centering
        \includegraphics[width=0.45\textwidth,height=0.4\textwidth]{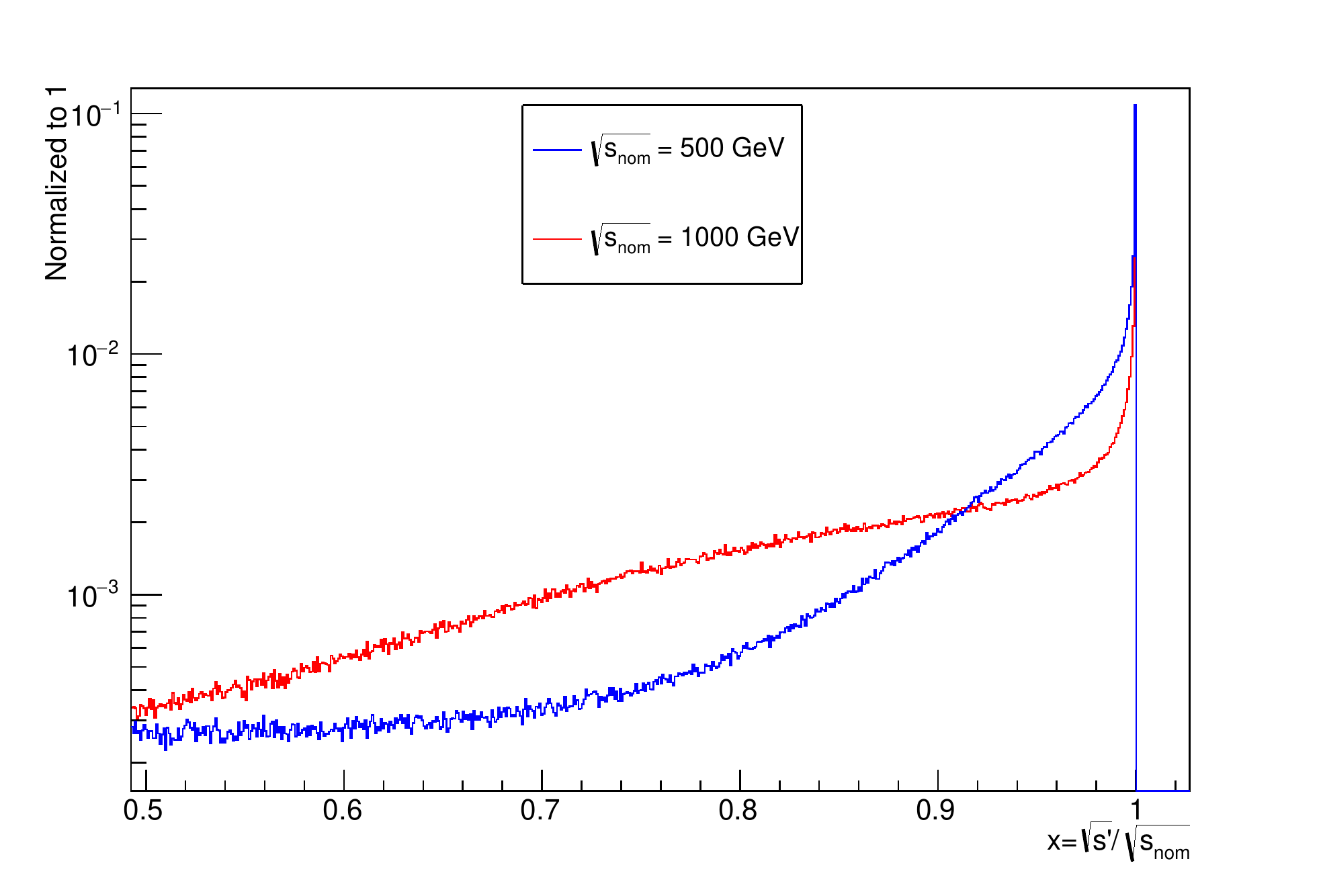}
        \caption{Distribution of the effective center of mass energy ($\sqrt{s'}$) including both beamstrahlung and ISR effects for the two cases of $\sqrt{s}=500$ and 1000 GeV. $\sqrt{s_{\textnormal{nom}}}$ is the nominal center of mass energy of the collider.}
\label{bs}
\end{figure}

\begin{figure}[h]
    \centering
        \includegraphics[width=0.45\textwidth,height=0.4\textwidth]{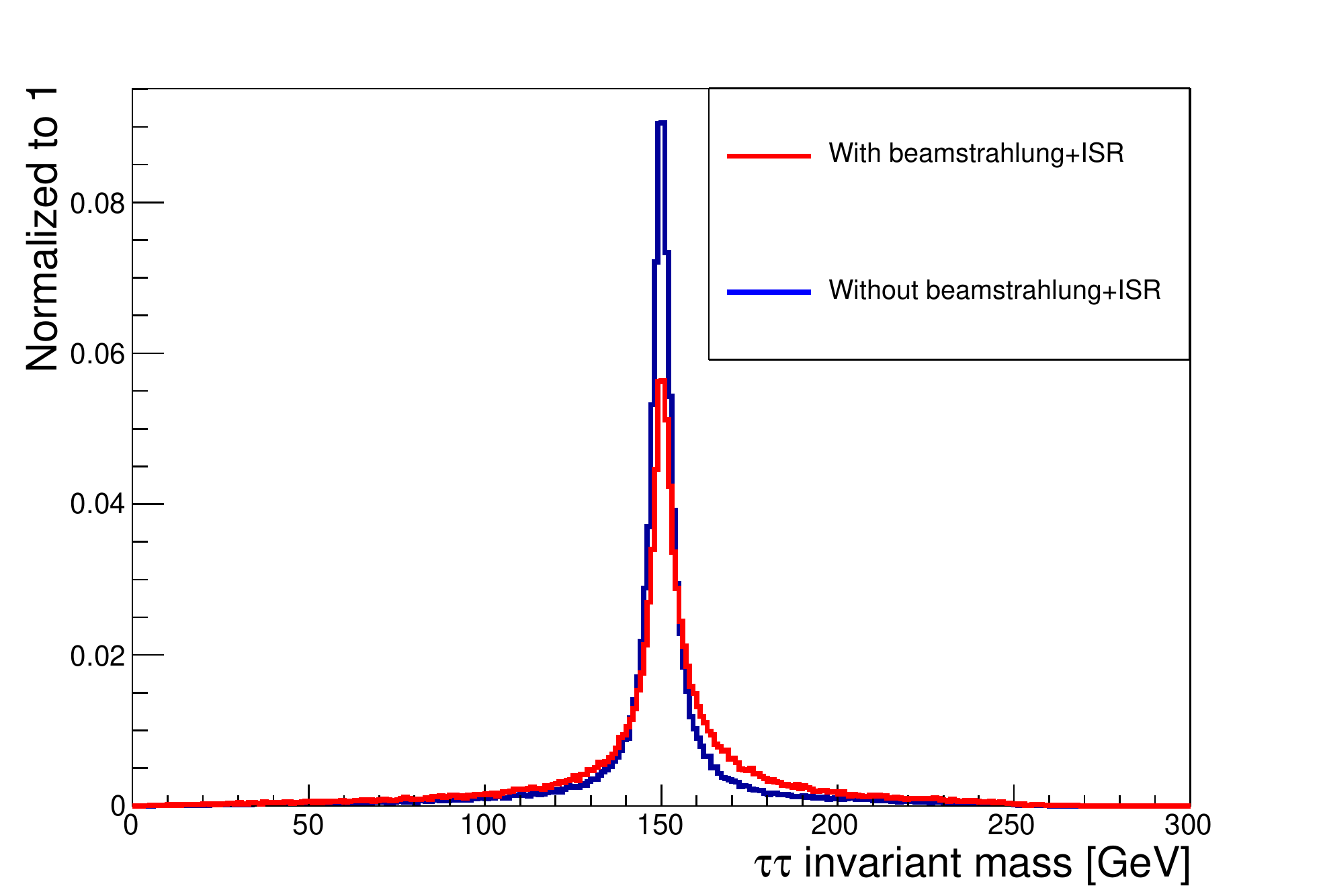}
        \caption{Signal of the $\tau$-jet pair invariant mass at $\sqrt{s}=500$ GeV with and without beamstrahlung+ISR.}
\label{smear}
    \end{figure}

In order to include initial state radiation (ISR) and the beam energy spectrum due to the effect of beamstrahlung, the event generation and detector simulation is performed in several steps. The hard scattering (the first few steps of each event before showering and hadronization) is performed by \texttt{CompHEP} \cite{comphep1, comphep2}. The beam parameters are given to \texttt{CompHEP} for each center of mass energy as shown in Tab. \ref{beam} \cite{ilctdrv3}. The electron spectrum due to ISR has been calculated in \cite{isr1}. In \texttt{CompHEP} a similar expression is used as obtained in \cite{isr2,isr3}. The effective energy spectrum of electron due to the effect of beamstrahlung is also simulated according to the function obtained in \cite{beamstrahlung}. The net effect is a small reduction in the effective center of mass energy as seen in Fig. \ref{bs} which is obtained by calculating the incoming $e^+e^-$ invariant mass, event by event, with or without ISR+beamstrahlung.  The distributions shown in Fig. \ref{bs} are in good agreement with the official ILC results \cite{lumispec,ilcrdr_physics}. The intrinsic beam spread is negligible compared to the ISR effect as shown in \cite{ilcrdr_physics}. The final results are therefore marginally affected as seen in Fig. \ref{smear} for a reconstructed $\tau$ jet pair invariant mass distribution of the signal. 

The output of \texttt{CompHEP} is stored in event files in \texttt{LHA} format \cite{lha} and is passed to \texttt{PYTHIA} for multi-particle interaction, final state showering and hadronization. 

Events generated by \texttt{PYTHIA} are then used by \texttt{DELPHES 3.4} \cite{delphes} for detector simulation with a detector card specialized for SiD detector at ILC \cite{dsid}. 

The jet reconstruction is performed by \texttt{FASTJET 3.1} with inclusive $k_t$ algorithm and jet cone size of 0.5 \cite{fastjet1,fastjet2}. The $k_T$ algorithm has been found to give better performance by reducing the clustering of low $p_T$ background in the forward/backward region \cite{HiggsCLIC}. The low $p_T$ events of $\gamma\gamma \to \textnormal{hadrons}$ are not simulated in this analysis. However, kinematic cuts $p_T > 5$ GeV and $15^{\circ} < \theta_{\tau}< 165^{\circ}$ are applied to all four jets in the event. Here $\theta_{\tau}$ is the $\tau$ jet polar angle with respect to the beam axis. 

A detailed study of $H \to \tau\tau$ and the beamstrahlung induced background from $\gamma\gamma$ and $e^{\pm}\gamma$ collisions shows that their contribution can be up to 50$\%$ of the SM background from direct $e^+e^-$ interaction at $\sqrt{s}=3$ TeV (CLIC) and 23$\%$ at $\sqrt{s}=1.4$ TeV \cite{HiggsCLIC}. Such background processes are very important at high center of mass energies (CLIC environment) due to increasing number of beamstrahlung photons. The corresponding contribution at ILC environment is expected to be smaller. However, a detailed study is needed for a realistic estimation of this overlay background and could be subject of a future study which is probably beyond the scope of \texttt{DELPHES} capabilities and needs full simulation of the detector response.      
\begin{table}[h!]
\centering
\begin{tabular}{ccc}
Beam parameters & $\sqrt{s}=500$ GeV, & 1000 GeV \\
\midrule
No. of particles / bunch ($\times 10^{10}$) & 2 & 1.74 \\
\midrule
RMS bunch length (mm) & 0.3 & 0.225 \\
\midrule
RMS horizontal beam size (nm) & 474 & 335 \\
\midrule
RMS vertical beam size (nm) & 5.9 & 2.7 \\
\midrule
\end{tabular}
\caption{Beam parameters for beamstrahlung simulation taken from Tab. 8.2 of ILC technical design report volume 3 (Accelerator baseline) \cite{ilctdrv3}. \label{beam}}
\end{table} 

Tables \ref{bp500} and \ref{bp1000} show the selected points in the physical mass basis for the two center of mass energies of $\sqrt{s}=500$  and 1000 GeV. The Higgs bosons are assumed to satisfy $m_A=m_{H^{\pm}}$ to ensure that deviation from SM in terms of $\Delta \rho$ is small enough and consistent with experimental value \cite{drho}. The selected points keep the Higgs bosons mass difference to be below the $Z$ boson mass to suppress $A \to ZH$ decay. Table \ref{b5001000} shows the corresponding background cross sections at $\sqrt{s}=500$ and 1000 GeV.

\begin{table}[h!]
\centering
\begin{tabular}{ccccc}
\midrule
\multicolumn{5}{c}{$\sqrt{s}=500$ GeV} \\
\midrule
 & BP1 & BP2 & BP3 &BP4\\
\midrule
$m_h$ & \multicolumn{4}{c}{125} \\
\midrule
$m_H$ & 150 & 150 & 200 & 200 \\
\midrule
$m_A$ & 150 & 200 & 200 & 250 \\
\midrule
$\tan\beta$ & \multicolumn{4}{c}{20}\\
\midrule
$\sigma$ $[fb]$	& 26.2 & 17.53 & 9.9 & 3.4  \\
\midrule
\end{tabular}
\caption{Signal benchmark points and their cross sections at $\sqrt{s}=500$ GeV.}
\label{bp500}
\begin{tabular}{ccccccc}
\\
\\
\midrule
\multicolumn{7}{c}{$\sqrt{s}=1000$ GeV} \\
\midrule
 & BP1 & BP2 & BP3 & BP4 & BP5 & BP6 \\
\midrule
$m_h$ & \multicolumn{6}{c}{125} \\
\midrule
$m_H$ & 150 & 200 & 200 & 250 & 250 & 300 \\
\midrule
$m_A$ & 150 & 200 & 250 & 250 & 300 & 300 \\
\midrule
$\tan\beta$ & \multicolumn{6}{c}{20}\\
\midrule
$\sigma$ $[fb]$	& 12.2 & 10.3 & 9.2 & 8.3 & 7.2 & 6.3 \\
\midrule
\end{tabular}
\caption{Signal benchmark points and their cross sections at $\sqrt{s}=1000$ GeV.}
\label{bp1000}
\begin{tabular}{ccccc}
\\
\\
\midrule
\small{Channel}&\small{$Z/\gamma*$}&\small{$ZZ$}&\small{$WW$}&\small{$t\bar{t}$}\\
\midrule
\multicolumn{5}{c}{$\sqrt{s}=500$ GeV}\\
\midrule
$\sigma$ $[fb]$  & \small{16830} & \small{582} & \small{7855} & \small{598} \\
\midrule
\midrule
\multicolumn{5}{c}{$\sqrt{s}=1000$ GeV}\\
\midrule
$\sigma$ $[fb]$ & \small{4318} & \small{234} & \small{3176} & \small{211} \\
\bottomrule
\end{tabular}
\caption {Considered background processes and their cross sections at $\sqrt{s}=500$ and 1000 GeV.}
\label{b5001000}
\end{table}
\section{Signal selection and analysis}
The \texttt{DELPHES} output is stored in \texttt{ROOT} files \cite{root} which contain reconstructed physical objects like electrons, muons and jets with additional flags for $b$-tagging and $\tau$-tagging results. The latter is what is used in the current analysis. 

The event selection starts from accessing reconstructed jets which pass kinematic requirements mentioned earlier. The $\tau$-tagging is in general based on a sophisticated algorithm. The $\tau$-jet should proceed in a narrow cone (isolated jet) accommodating one or three charged tracks associated with one prong or three prong hadronic decay with the hardest track carrying a large fraction of the jet energy. The $\tau$-tagging algorithm which is used in this analysis relies on \texttt{DELPHES} implementation and is based on a simple matching method, where, the $\tau$ particle four-momentum is calculated from its decay products using generator level information. Then for every jet from the \texttt{FASTJET} output, a search is performed between all $\tau$ particles to find the closest $\tau$ to the jet under study. If $\Delta R$ between the $\tau$-jet and the $\tau$ particle from generator level information is less than 0.2, the jet is identified as a $\tau$-jet with an efficiency of  $90\%$ which is the case for $\tau \to \pi\nu,~\rho\nu$ and $a_1\nu$ (3-prong decay) (see Tab. III-6.3 of ILC TDR \cite{ilctdrv4}). Here $\Delta R$ is defined as $\Delta R=\sqrt{(\Delta \eta)^2 + (\Delta \phi)^2}$ with $\eta=-ln\tan(\theta/2)$. The angles $\theta$ and $\phi$ follow the standard definitions: the polar and azimuthal angles. 

The $\tau$ jet mistagging rate is also taken into account using the matching method with efficiency of 0.1$\%$. 

Events with at least four identified $\tau$-jets are selected for the analysis. Requiring two $\tau$-jets allows for a large background from single $Z$ boson production. 

For a reasonable event reconstruction, a $\tau$ four-momentum correction is performed \cite{MSSMHiggsLEP, Kanemura2}. The idea is based on the fact that the number of momentum and energy conservation equations are equal to the number of scaling factors to be applied on the four $\tau$ jets in the event. Therefore one can solve the four simultaneous equations (three for the total momentum conservation and one for the total energy conservation) and find the four unknown factors as in Eq. \ref{zfactors} (The unknown factors are $z_1,~z_2,~z_3,~z_4$ and momentum/energy numerical indices denote the four $\tau$ jets $\tau_1,~\tau_2,~\tau_3,~\tau_4$). This is of course based on the assumption that the $\tau$ jet direction has correctly been measured and a common factor can be applied to all its four-momentum components. This assumption is based on collinear approximation which implies that the visible $\tau$ jet and the associated neutrino are collinear in the high energy limit and the jet flight direction is approximately that of the $\tau$ lepton. Applying this correction, a dramatic improvement is obtained in the $\tau$ jet pair invariant mass distribution. 
\begin{align}
\begin{split}
z_1~p_1^x + z_2~p_2^x + z_3~p_3^x + z_4~p_4^x &= 0, \\
z_1~p_1^y + z_2~p_2^y + z_3~p_3^y + z_4~p_4^y &= 0, \\
z_1~p_1^z + z_2~p_2^z + z_3~p_3^z + z_4~p_4^z &= 0, \\
z_1~E_1 + z_2~E_2 + z_3~E_3 + z_4~E_4 &= \sqrt{s}\\
\end{split}
\label{zfactors}
\end{align}
It should be noted that the four factors obtained in this way are required to be positive which is the case for most signal events but is useful to reduce the fake rate from SM background. When the $\tau$ jet four-momenta are rescaled, they are sorted in terms of the new energies and then the pairing is performed.

There are two ways to find the right $\tau$-jet pairs from the Higgs bosons decays. One way is to find the $\tau$-jet pair with minimum $\Delta R$. This approach has shortcomings due to the fact that with increasing Higgs bosons masses, heavy Higgs bosons with sum of their masses being close to the center of mass energy, tend to be produced almost at rest in the laboratory frame and $\Delta R$ between their decay products tends to be large. Therefore the algorithm of finding the $\tau$ jet pair with minimum $\Delta R$ starts to be less efficient with increasing Higgs bosons masses. 

Instead, the correct pairs of the $\tau$ jets are found by sorting them according to their energies and defining the two $\tau$ jets with maximum and minimum energies as one pair and the other two as the second pair. Therefore if $\tau$ jets are sorted in terms of their energies and labeled as $\tau_1,~\tau_2,~\tau_3,~\tau_4$, the two pairs are made of $\tau_1,\tau_4$ and $\tau_2,\tau_3$. 

Each one of the above pairs may come from $H$ or $A$. Therefore invariant mass distributions of both $\tau_1,\tau_4$ and $\tau_2,\tau_3$ pairs show the two Higgs bosons peaks if the Higgs bosons masses are different. This is a unique feature of this production channel.  

Tables \ref{eff500} and \ref{eff1000} show the signal and background selection efficiencies in the two scenarios of $\sqrt{s}=500$ and 1000 GeV. No SM background other than $ZZ$ survives at the end and the signal to background ratio is large in all cases. The high signal significance which is due to the small cross section of the background and reasonable selection efficiencies, reveals that the signal can be observed earlier before the scaled integrated luminosity of 500 $fb^{-1}$ is achieved. 

For comparison, the current results lead to 1015 signal and 2 background events at 500 $fb^{-1}$ for the first benchmark point which is $m_H=m_A=150$ GeV and is the closest to the point chosen by \cite{Kanemura2}. This means that at 100 $fb^{-1}$ one would have ~203 signal vs 0.4 background events which is consistent with their results.

\begin{figure*}[h]
    \centering
        \includegraphics[width=0.9\textwidth,height=0.8\textwidth]{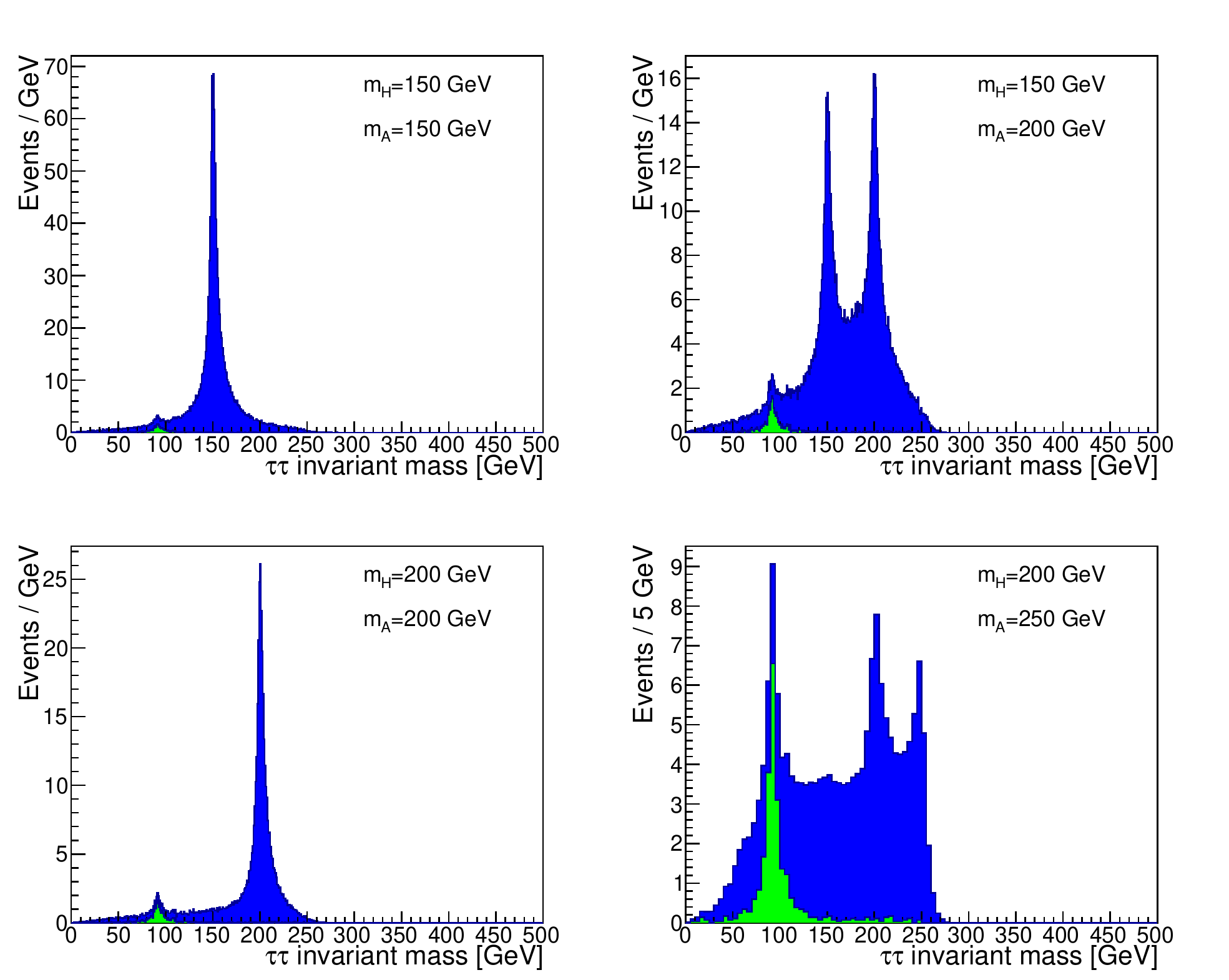}
        \caption{Signal of the $\tau$-jet pair invariant mass in different benchmark points at $\sqrt{s}=500$ GeV}
\label{s1}
    \end{figure*}

\begin{figure*}[h!]
    \centering
        \includegraphics[width=0.9\textwidth,height=1.2\textwidth]{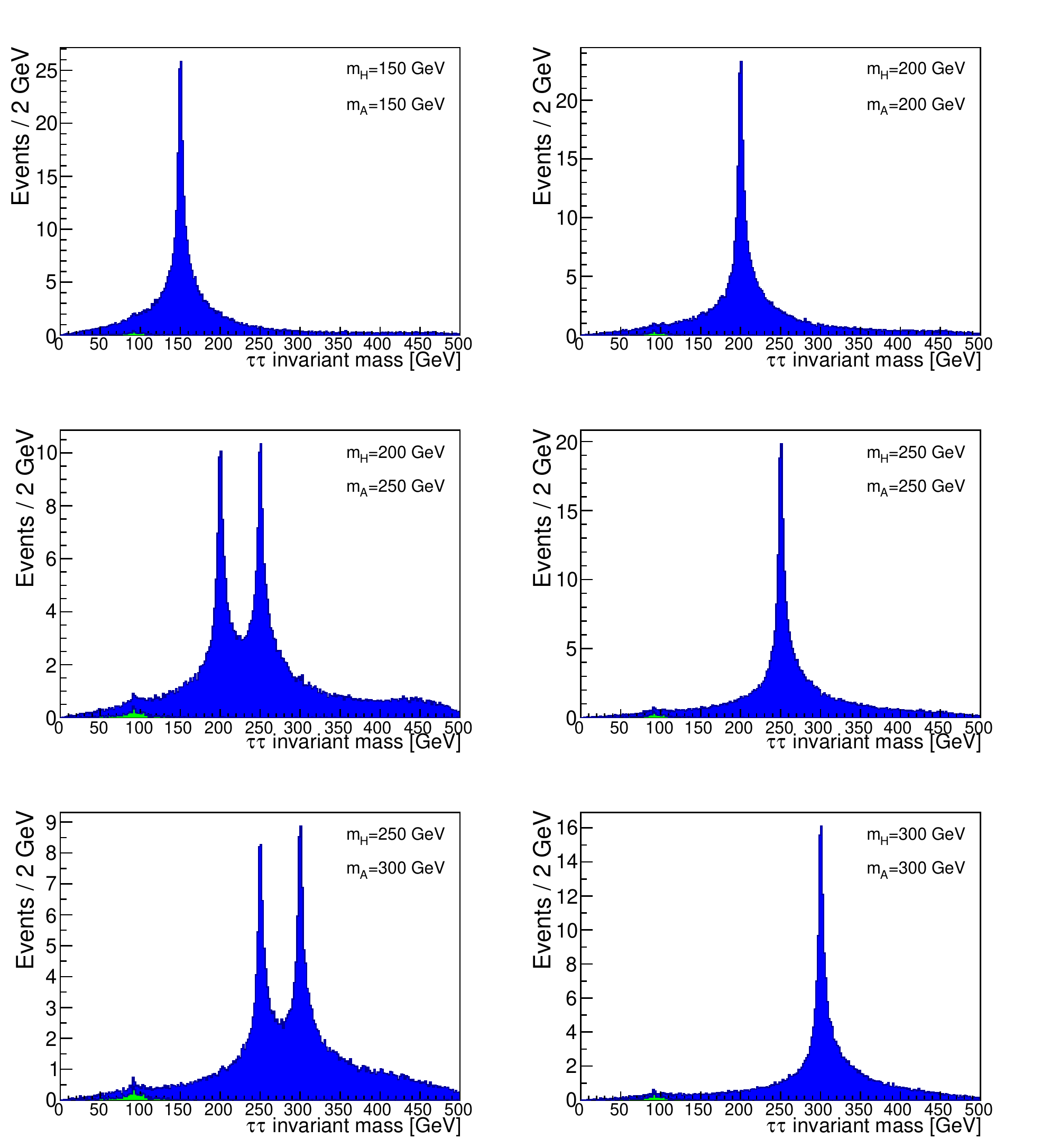}
        \caption{Signal of the $\tau$-jet pair invariant mass in different benchmark points at $\sqrt{s}=1000$ GeV}
\label{s2}
    \end{figure*}

\begin{table}
\begin{tabular}{ccccc}
\\
\\
\midrule
\multicolumn{5}{c}{$\sqrt{s}=500$ GeV, int. lumi = 500 $fb^{-1}$}\\
\midrule
\multicolumn{5}{c}{Background processes}\\
\midrule
 & ZZ & $t\bar{t}$ & WW & $Z/\gamma$ \\
\midrule
$\epsilon_B$ & $~7\times 10^{-6}~$ & $~<10^{-7}~$ & $~<10^{-7}~$ &$~<10^{-7}~$ \\ 
\midrule
\multicolumn{5}{c}{Signal processes}\\
\midrule
 & BP1 & BP2 & BP3 &BP4\\
\midrule
$\epsilon_S$ & 0.077 & 0.076& 0.076 & 0.064 \\ 
\midrule
S & 1015 & 670 &381  & 109 \\
\midrule
B & \multicolumn{4}{c}{2}\\
\midrule
S/B &482  & 318 &180 & 47 \\
\midrule
$S/\sqrt{B}$ & 700 & 462 & 262 & 72 \\
\midrule
$S'$ & 103& 80 & 57 & 25\\
\midrule
\bottomrule
\end{tabular}
\caption {Signal and Background selection efficiencies and the signal significance in different benchmark points at $\sqrt{s}=500$ GeV. The second significance estimator $S'$ is defined as $S'=\sqrt{2((S+B)\ln(1+S/B)-S)}$ (\cite{Kanemura2, Kanemura3}). Background efficiencies have been obtained from a sample of 10M events and no event of $t\bar{t}$, $WW$ or $Z/\gamma$ remains.} 
\label{eff500}
\end{table}
\begin{table}
\begin{tabular}{ccccccc}
\\
\\
\midrule
\multicolumn{7}{c}{$\sqrt{s}=1000$ GeV, int. lumi = 500 $fb^{-1}$}\\
\midrule
\multicolumn{7}{c}{Background processes}\\
\midrule
 & ZZ & $t\bar{t}$ & WW & $Z/\gamma$ \\
\midrule
$\epsilon_B$ & $~4\times 10^{-6}~$ & $~<10^{-7}~$ & $~<10^{-7}~$ &$~<10^{-7}~$ \\ 
\midrule
\multicolumn{7}{c}{Signal processes}\\
\midrule
 & BP1 & BP2 & BP3 &BP4 & BP5 & BP6\\
\midrule
$\epsilon_S$ & 0.053 & 0.066& 0.07 & 0.073 &  0.075 & 0.076\\ 
\midrule
S &  328 & 342 & 322  & 304 & 272 & 241 \\
\midrule
B & \multicolumn{6}{c}{0.2}\\
\midrule
S/B & 652 & 680 & 641 & 605 & 540 & 480\\
\midrule
$S/\sqrt{B}$ & 462 & 482 & 455 & 428 &  383 & 340 \\
\midrule
$S'$             & 60  & 61   & 59  & 57 &  54 & 50 \\
\midrule
\bottomrule
\end{tabular}
\caption {Signal and Background selection efficiencies and the signal significance in different benchmark points at $\sqrt{s}=1000$ GeV. $S'$ is defined as mentioned in the caption of Tab. \ref{eff500}.}
\label{eff1000}
\end{table}
Figures \ref{s1} and \ref{s2} show the signal on top of the background. The size of the MC sample used to obtain these plots is two orders of magnitude larger than the actual size of the expected sample from the real experiment. However the final results have been scaled to the integrated luminosity of 500 $fb^{-1}$. The detector effects have been propagated into the final results in terms of momentum and energy smearing. As seen from these figures, the signal is well visible in a clean environment with small background contamination. The signal with $m_H=200$ GeV and $m_A=250$ GeV has a small cross section at $\sqrt{s}=500$ GeV but is still observable. The analysis at $\sqrt{s}=1$ TeV is effectively a background free analysis with a very tiny background at the $Z$ mass. Signal of the first benchmark point is expected to be visible at even lower center of mass energy corresponding to the ILC operation at $\sqrt{s}=350$ GeV. This expectation can be verified in detail but what was shown in this analysis is the dramatic potential of linear colliders for the 2HDM Higgs boson observation at moderate masses even though at a specific model type.  
\section{Conclusions}
Signals of the 2HDM Higgs bosons were analyzed for the case of unpolarized $e^+e^-$ collisions at $\sqrt{s}=500$ and 1000 GeV. Different benchmark points were studied for the two operation scenarios. Assuming 2HDM type IV as the theoretical framework, the Higgs bosons decay to $\tau\tau$ was analyzed focusing on the hadronic $\tau$ decays. The detector simulation was performed based on parameters from the SiD detector studies at ILC. Results can be summarized as follows. 

The Higgs boson decay to $\tau\tau$ is dominant as long as decays to gauge bosons are kinematically forbidden. The decay to $\tau\tau$ reaches $\sim 80\%$ at $\tan\beta=3$ for both $H$ and $A$. The signal has a small sensitivity to the value of $\tan\beta$ and results are valid for $\tan\beta>3$ with negligible variation. There is also negligible sensitivity of BR($H/A \to \tau\tau$) to the Higgs bosons masses as long as $m_A-m_H < m_Z$. The signal rate thus depends only on phase space (Higgs bosons masses) and tends to decrease when $m_H+m_A$ reaches the collider center of mass energy. This can be seen by following results of the analyses of the chosen benchmark points. However, due to the reasonable performance of kinematic algorithm used in the analysis, striking signals are observable at high statistical significances even when $m_H+m_A$ is below $\sqrt{s}$ by 50 GeV (BP4). The signal is thus expected to be observable for all $m_H$ and $m_A$ values as long as $m_H+m_A$ is less than $\sqrt{s}$. The conclusion does not depend sizably on the value of $\tan\beta$. The value of $\alpha$ should of course be extracted from $\sin(\beta-\alpha)=1$ which is the SM-like assumption. 

Results obtained in the current analysis are beyond the reach of LHC as their main search channel is through $A \to ZH$ decay which is suitable for $\tan\beta<3$ and large mass splitting between $H$ and $A$. At the ILC, The signals of the Higgs bosons are observable as sharp distributions on top of the background at integrated luminosity of 500 $fb^{-1}$ and already much lower luminosities are sufficient for discovery. Therefore the signal cross section can be measured from data with a reasonable precision. Such precision measurements can be used to verify the model type (through Higgs boson coupling to leptons) and $\alpha$ and $\beta$ angles through Higgs boson coupling to gauge bosons. Therefore it is expected that such a background free signal would serve as a tool to analyze the 2HDM parameter space.  
\section*{Acknowledgments}
We would like to thank the college of sciences at Shiraz university for providing computational facilities during the research program.
%
%
\end{document}